\documentclass[runningheads]{llncs}
\usepackage{graphicx}
\usepackage{appendix}
\usepackage{siunitx}
\usepackage{caption}
\usepackage{subcaption}
\begin{document}
\title{Analyzing the Evolution of Inter-package Dependencies in Operating Systems: A Case Study of Ubuntu}
%
%
\author{Victor Prokhorenko\inst{1,2} \and Chadni Islam\inst{3} \and Muhammad Ali Babar\inst{1,2}}

\authorrunning{V. Prokhorenko et al.}

\institute{CREST - The Centre for Research on Engineering Software Technologies, the University of Adelaide, Australia 
\\ \email{{victor.prokhorenko, ali.babar}@adelaide.edu.au}\\ \and
Cyber Security Cooperative Research Centre (CSCRC), Australia\\ 
\and Queensland University of Technology, Brisbane, Australia 
\email{chadni.islam@qut.edu.au}}

\maketitle              
\begin{abstract}
\vspace{-10pt}
An Operating System (OS) combines multiple interdependent software packages, which usually have their own independently developed architectures. When a multitude of independent packages are placed together in an OS, an implicit inter-package architecture is formed. For an evolutionary effort, designers/developers of OS can greatly benefit from fully understanding the system-wide dependency focused on individual files, specifically executable files, and dynamically loadable libraries. We propose a framework, DepEx, aimed at discovering the detailed package relations at the level of individual binary files and their associated evolutionary changes. We demonstrate the utility of DepEx by systematically investigating the evolution of a large-scale Open Source OS, Ubuntu. DepEx enabled us to systematically acquire and analyze the dependencies in different versions of Ubuntu released between 2005 (5.04) to 2023 (23.04). Our analysis revealed various evolutionary trends in package management and their implications based on the analysis of the 84 consecutive versions available for download (these include beta versions). This study has enabled us to assert that DepEx can provide researchers and practitioners with a better understanding of the implicit software dependencies in order to improve the stability, performance, and functionality of their software as well as to reduce the risk of issues arising during maintenance, updating, or migration.

\end{abstract}

\textit{This work is accepted for publication in The 17th European Conference on Software Architecture (ECSA 2023), 
Istanbul, Turkey.}

\section{Introduction}

Combining multiple independent software packages together is commonly used to form complex inter-connected ecosystems. A typical example of such large software ecosystems is various Linux distributions. Such ecosystems tend to consist of hundreds or thousands of packages, libraries, binaries, and configuration files with an order of magnitude more dependencies among them~\cite{unix_evolution_TSC}, \cite{unix_44}. 

Developers and researchers have expressed interest in software complexity measurement in an attempt to reason about characteristics of large code bases \cite{MetricsFaults}. Software complexity is viewed as a result of different design decisions and implementation specifics and is a crucial component of long-term effects like the maintainability of software~\cite{softwareComplexity}. Although software complexity is a crucial consideration for package managers, Linux distributors, and maintainers, we currently have limited knowledge about the evolution of this complexity over the software lifespan. While the complexity of individual packages is tamed by their corresponding developers, combining thousands of packages materializes a new emergent layer of complexity. It is also uncertain whether different metrics for measuring software complexity exhibit similar or varying patterns of evolution.

A significant amount of research has extensively explored source-level software complexity~\cite{SoftwareMetricsSurvey}. As a result, various complexity metrics have been defined, such as cyclomatic, branching, or data flow complexity~\cite{SoftwareMetrics}. These metrics are primarily used for software design, debugging, and optimization purposes \cite{softwareComplexity}. 

These metrics are, however, not applicable when analyzing closed-source software distributed only in binary form without access to the source code. In such cases, binary dependency analysis is required to understand the interactions and dependencies between compiled binary executables. Additionally, even when source code is available, there may be situations where the compiled binary may behave differently from what is expected due to specific environment configurations. Thus, binary dependency analysis can provide a more accurate and complete understanding of run-time software behavior, which can be crucial for identifying potential issues or vulnerabilities.

This work considers an OS as a whole rather than focusing on analyzing individual software binaries. Considering an OS enables the identification of cross-application relations, which make up an emergent \textbf{inter-package relation architecture} instead of just the intra-package software complexity. We propose a framework that enables the extraction of binary-to-library dependencies and constructs a full OS dependency graph to obtain insights on overall OS complexity which we determine through inter-package dependency coupling. By coupling we mean any type of dependency of one code fragment on another (library inclusion, function call, etc).

Our study focused on Ubuntu as a case study to examine the evolution of large software ecosystems over almost two decades. Through empirical research and evidence-based findings, we aimed to assess the current state of package, library, and binary dependencies and identify areas for improvement in management tools, policies, and ecosystem analysis platforms. 
We believe that a deep understanding of emergent inter-package architecture resulting from combining a multitude of independently developed software subsystems would benefit software developers and OS maintainers. The proposed techniques and tools are expected to minimize manual labor associated with multi-package maintenance.

Following are the key contributions of our work
\vspace{-5pt}
\begin{itemize}
    \item We have introduced a framework for dependency coupling analysis for multi-package software to extract the inter-package relations architecture that is applicable to a broader range of OS due to the binary-level analysis.
    \item We have defined four techniques to quantitatively measure software coupling in terms of executable and dynamically loadable library dependencies at different granularities.
    \item We have investigated the evolution of Ubuntu OS in terms of the proposed library presence dependency type, which revealed the changes in OS-wide inter-package relations over time.
\end{itemize}

\vspace{-12pt}

\section{Background and Motivation}
\vspace{-5pt}
\subsection{Software Complexity}

Throughout the lifetime of any software system, various code modifications must be implemented in order to adapt to ever-changing user requirements and environmental conditions. An intuitive expectation is that large and complex software systems may be more difficult to update and maintain. Thus, in efforts to gain a stricter definition of complexity, multiple code complexity measurement techniques, such as straightforward line count or cyclomatic complexity, have been proposed so far \cite{SoftwareMetrics}. However, analyzing multiple diverse software systems as a whole is not trivial due to (i) lack of access to the source code of all third-party components,
    (ii) lack of formal interoperability specification and 
    (iii) highly dynamic state of execution environment at run time.

Several techniques are typically employed to handle the growing complexity of large software systems (such as a full OS). For instance, the system package manager may track package dependency information at the OS level. This tracking enables detecting incompatibilities between separate software subsystems and repairing them if possible. Unfortunately, manual labor is commonly used in determining and maintaining information on such version-level incompatibilities \cite{PackageConflict}. Due to the large number of files in a typical OS, manual efforts typically target only high-level dependency definitions, such as package level only~\cite{LinuxPackage_IEEE}. As each package may consist of multiple files containing executable code (i.e., executable binaries and libraries), such package dependency understanding may not represent the dependencies precisely.

Further challenges arise due to modern complex software systems commonly developed in various programming languages. For instance, purely-binary compiled languages are intertwined with interpreted script languages leading to execution flow frequently being transferred between them. The dependency chains within such complex systems may propagate through a significant portion of files in the file system through the indirect reliance of different code fragments on each other. A typical example includes PHP web pages relying on the PHP interpreter, web server, and third-party PHP libraries. Such immediately obvious (direct) dependencies, in their turn, recursively rely on other system-provided and third-party libraries. Therefore we argue that automated and precise dependency tracking would benefit software system maintainers and administrators and may provide useful insight to software developers.

\vspace{-10pt}
\subsection{Code dependency types}

One piece of code can depend on another in numerous ways. For instance, within the source code analysis context, a function may call other functions. Similarly, class methods may work by invoking other class methods. These types of dependencies present in the same code base are well understood and routinely used in modern IDEs (Integrated Development Environments) to aid software developers. In contrast, cross-language code dependencies spanning across multiple independently developed software systems are less formal and challenging to identify. For instance, a PHP-oriented IDE would not detect incompatible changes in the \texttt{libc} library which is required by the PHP interpreter itself.

\begin{figure}
    \centering
    \includegraphics[scale=0.14]{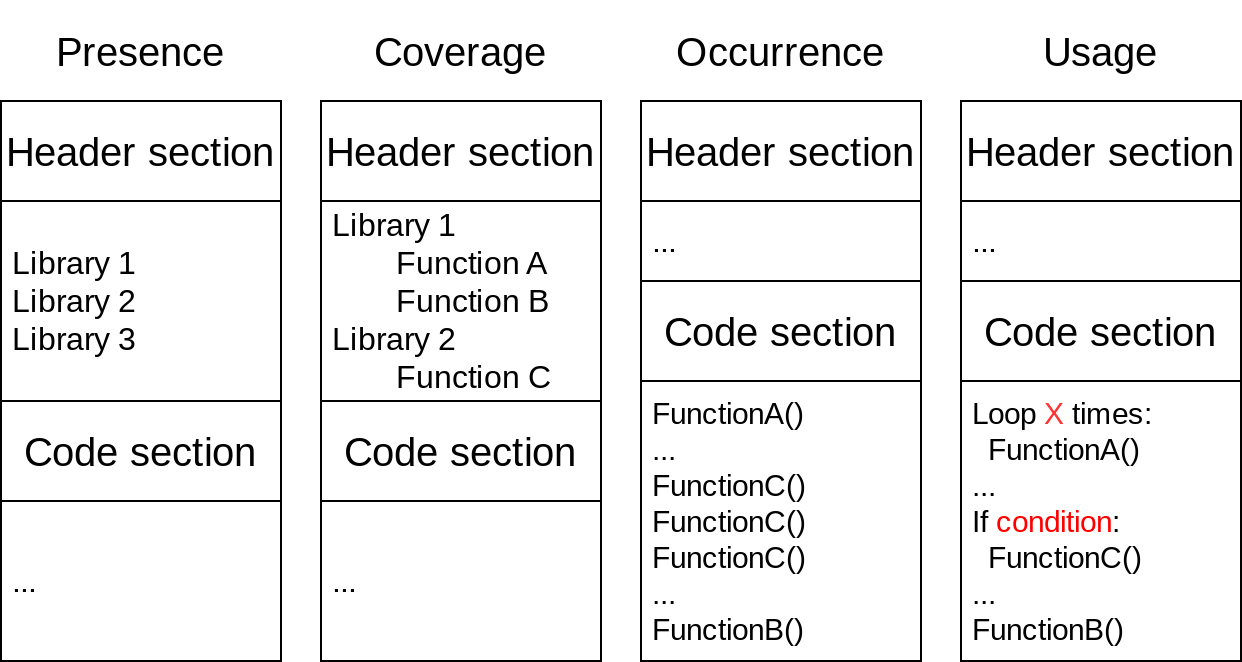}
    \caption{Executable code dependency measurement approaches}
    \label{fig:codeDepency}
    \vspace{-15pt}
\end{figure}

Focusing solely on software running within the same OS while not taking network-based dependencies into consideration, we propose the following four conceptual types of dependencies suitable in the executable code analysis context. These four types include (i) the presence of third-party libraries, (ii) the extent of library coverage, (iii) library function call occurrences, and (iv) the run-time usage of functions (Figure~\ref{fig:codeDepency}).

\vspace{-15pt}

The \textbf{third-party library presence dependency} relates to file-level granularity. This type of dependency indicates a requirement for a dynamically loadable library to be \textit{present} in the system for an executable binary to be able to load and start. In Windows-based systems, libraries and executables are denoted by .dll and .exe file extensions, while on Linux-based these are .so and typically extension-less ELF (Executable and Linkable File) correspondingly. While high-level, this file granularity is crucial as a missing library file typically causes the executable file loader to indicate an error and prevents any further file execution.

\textbf{Coverage dependency} focuses on the library fragments (e.g., functions or class methods) that a developer explicitly uses or relies on. This type of dependency refers to specific \textit{function existence requirements}. Thus, the library coverage aspect reflects the degree of reliance on a given library by the executable. Depending on the OS, programming language, and execution environment, individual function-level requirements can be implemented in various ways. For instance, in the context of the Windows PE executable, the list of required functions is tied to a specific library. In contrast, the lists of required libraries and functions are independent in the Linux ELF executable~\cite{ELFspec}. These implementation specific differences complicate coverage analysis in the general case.

\textbf{Function occurrence} dependency type attempts to provide further insight into the code dependency by observing that a single external function can be \textit{referred} to multiple times in the original code. For instance, some heavily used functions can be mentioned all over the code, while some rarely used functions may only appear once. Extracting this type of dependency is extremely complicated and involves computationally-heavy disassembling of compiled code or parsing of interpreted languages. Initial unoptimized attempts revealed a significant time overhead for extracting such occurrence-level dependencies. While certain optimizations can be taken for production-ready usage, it can be concluded that this type of analysis is currently unsuitable for real-time applications.

Lastly, \textbf{dependency usage} refers to the actual run-time external code flow control transfers (i.e., the actual function \textit{calls}). This level of detail may, for example, reveal that one function call is contained within a high-count loop while other function calls may be a part of a condition rarely satisfied at run time. Run-time observation would reveal a deeper understanding of the level of reliance on third-party libraries in both cases. Despite seemingly most accurate and closest to reality, relying on this type of dependency suffers from a major drawback. Different executions or instances of the same executable may exhibit different behavior due to different run-time conditions. In other words, observing a single execution does not guarantee to reveal all external code usage cases.

Note that a purposefully crafted executable may incorporate external dependencies that would not be reflected using the proposed dependency measurement techniques. For instance, if an executable downloads  code over the network and executes it in place, no third-party library references, function names, or function calls related to the downloaded code may be present in the original executable. Moreover, the downloaded code downloaded can be different on each program invocation, making any dependency analysis futile in such a context. Based on the identified dependency types, we propose an extensible plugin-based framework suitable to extract code dependencies for various types of executable code.

\vspace{-5pt}
\section{Our Approach and Implementation}

Analyzing the full file system enables a more complete and consistent understanding of the dependencies. Software developers only \textit{express a requirement} for dynamically loadable library \textit{presence}, but do not have actual guarantees of the library's \textit{existence} in a given system. We implement a Python-based proof of concept solution to analyze system-wide dependencies.


On a conceptual level, our proposed approach for \textbf{Dep}endency \textbf{Ex}traction (\textbf{DepEx} consists of a file system scanner, a plugin dispatcher, multiple user-definable file-type-specific plugins, and the resulting database. The following steps provide an overview of the DepEx operation:
\vspace{-5pt}
\begin{itemize}
    \item The existing dependency extraction plugins (also Python-based) are queried to prepare the list of all supported file types
    \item The specified file system is iterated over and each file of a supported type is passed to a corresponding plugin for dependency extraction
    \item The dependencies extracted by the plugin are stored in an SQLite database
\end{itemize}

Having the knowledge of individual file type structures, each plugin is responsible for external dependency detection and extraction. Note that while the current implementation assumes one-to-one relation between file types and plugins, it is possible for multiple plugins to process the same files to extract different types of dependencies. 
While we have implemented a proof of concept plugins for PHP, Bash, and, to a lesser degree, Python scripts, in this research we primarily focus on ELF executables and .so libraries with the library presence dependency.

Once the unattended phase of the dependency extraction is complete, several interactive analysis and usage scenarios become accessible. These include visualization, statistical reporting, and forward and reverse update impact estimation. For instance, various system health characteristics, such as "number of missing libraries" or "number of executables with unfulfilled dependencies" can be queried and plotted if necessary. Similarly, update impact calculation enables obtaining the list of executables and libraries that would be potentially affected in case a given library is updated.

\begin{figure}
     \centering
     \begin{subfigure}[b]{0.49\textwidth}
         \centering      \includegraphics[width=\textwidth]{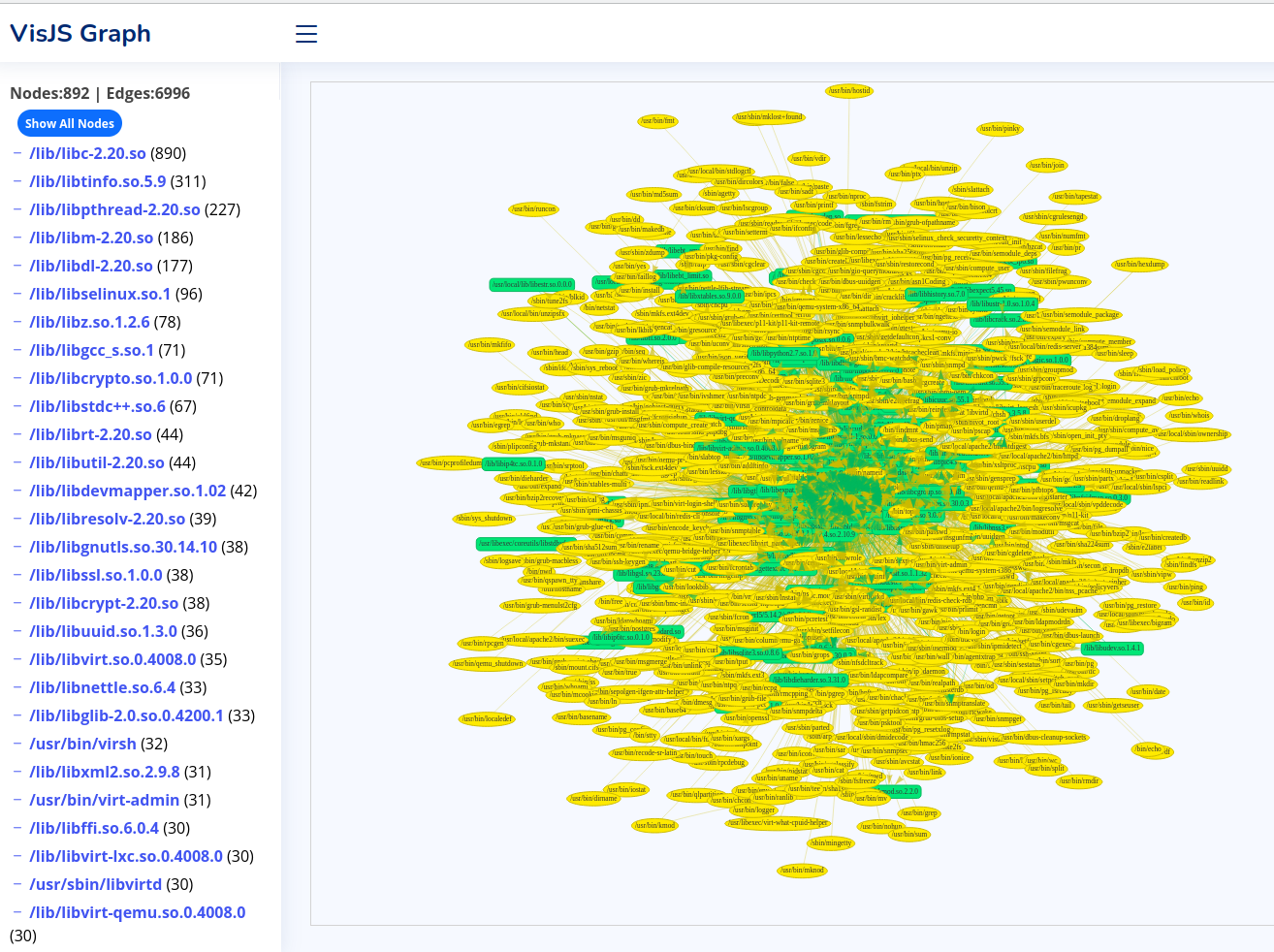}
    \caption{Full system view}
         \label{fig:directdep}
     \end{subfigure}
     \hfill
     \begin{subfigure}[b]{0.49\textwidth}
         \centering      \includegraphics[width=\textwidth]{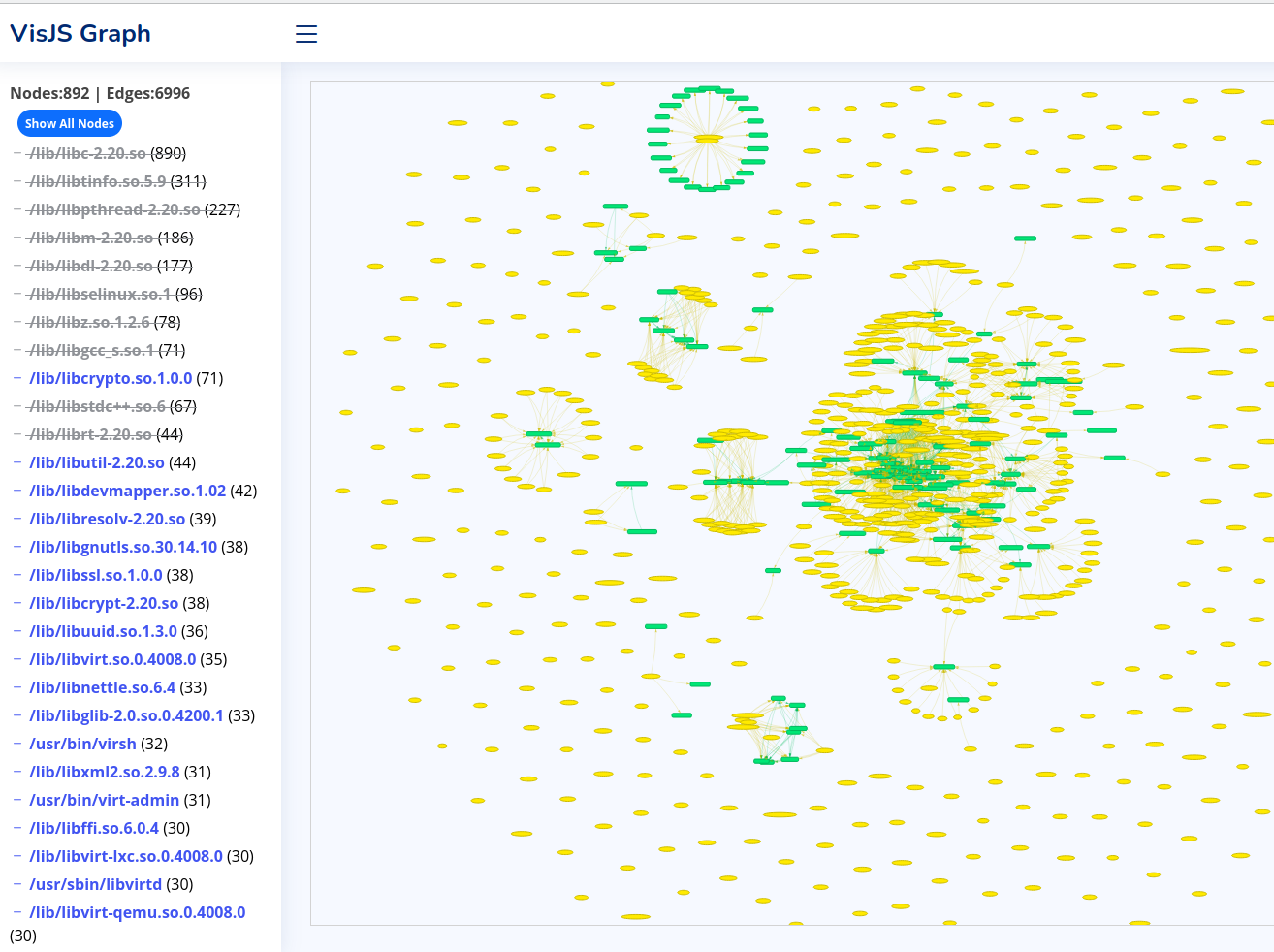}
    \caption{\small{Filtered system view}}
    \label{fig:visual}
     \end{subfigure}  
     \hfill
    \caption{Dependency visualization filtering effects}
    \label{fig:three graphs}
    \vspace{-15pt}
\end{figure}

In order to aid comprehension of the large amounts of data collected, we developed a visualization subsystem. Using DOT language for graph representation enables rendering the resulting graphs using existing tools as well (such as GraphViz or Gephi). While the individual executable file graphs were readable, the full-system dependency graph was too cluttered for human comprehension. At this stage, interactive filtering was implemented to allow the hiding of popular libraries responsible for most of the visual noise (as shown in Figure~\ref{fig:visual}). We are also planning to implement automated filtering based on various features, such as node type, sub-string matching, and popularity.

Other auxiliary scripts for dependency graphs exploration include querying all binaries and libraries that depend on a given library (\texttt{who-uses}) and individual binary/library dependency graph generation (\texttt{get-deps} and \texttt{get-all-deps}). Individual library dependencies can also be visualized in a more detailed view.

\vspace{-10pt}

\section{Studying the Architectural Aspects of Ubuntu}
We focus on the following Research Questions (RQs) to investigate the file-level package relation architecture in Ubuntu systems using DepEx. We considered the \textit{presence} dependency in this case study. We collected and analyzed the dependencies of 84 consecutive live Ubuntu Linux images that span over 18 years of development and evolution. The research questions we primarily focus on revolve around the emergent inter-package OS-wide architecture implicitly \textit{forming as a result of combining} multiple independent software packages as well as the related \textit{architectural changes observed} throughout longer time periods. In addition, we investigate the \textit{complexity perception} from the perspectives of individual software package developers and whole system maintainers.
\vspace{-5pt}
\begin{itemize}
    \item RQ1. How do binary-to-library dependencies manifest in the Ubuntu OS in terms of a system-wide dependency graph?


    \item RQ2. What is the difference between individual library complexity directly exposed to developers vs. overall internal system complexity that emerges as a result of combining multiple subsystems together (direct vs. recursive dependencies)?
    
    \item RQ3. How does the whole Ubuntu OS binary-to-library dependency graph evolve over a longer period?
\end{itemize}

Having high popularity, rich history, and open-source nature, Ubuntu serves as a comprehensive data source. Despite other Linux distributions, such as Alpine, gaining popularity, we were unable to find another dataset comparable in size and quality. Specifically, older Alpine versions were unavailable for download and Debian produced fewer live images.

Throughout the development of our DepEx framework, we relied on well-established existing open-source software, such as 
squashfs-tools\footnote{\url{https://github.com/plougher/squashfs-tools}}, binutils\footnote{\url{https://www.gnu.org/software/binutils/}} and ldd\footnote{\url{https://man7.org/linux/man-pages/man1/ldd.1.html}}. SquashFS-related tools were used to expose compressed live Ubuntu images for analysis. Note that different versions of \texttt{SquashFS} had to be used depending on the age of the Ubuntu image. Binutils package, particularly the GNU \texttt{nm} tool, was used to extract ELF-specific data such as imported library names. Lastly, \texttt{ldd} was used to extract library search locations. Special precautions had to be taken to lookup for the library paths inside the mounted image rather than resolving paths within the host system that conducted the analysis. For this purpose, we relied on standard Linux \texttt{chroot} functionality.

Solely mounting the Ubuntu ISO files directly does not provide access to the live file system, as another layer of compression is typically present for disk space optimization purposes. Thus, we implemented a two-step unpacking process to gain visibility of the inner live file system.

Interestingly, extracting the images generated over 18 years revealed how \textit{\textbf{}live image preparation} changed over time. We noticed different compression techniques used throughout the time period analyzed that ranged from compressed \textit{loop files} (\textit{cloop}) to SquashFS versions 2.1-4.0. We also observed that modern SquashFS kernel modules could not transparently mount images compressed by older versions. Thus, we developed a supporting script to provide access to all of the downloaded images in a uniform manner.

Using our DepEx framework, we recursively built the full library dependency graph for each identified executable using \texttt{readelf}, \texttt{nm} and \texttt{ldconfig} tools. Extracting library dependencies requires analyzing \texttt{RPATH} and \texttt{RUNPATH} variables, system library cache as well as the binary executable file path. Finally, we used an SQLite database to store the collected dependency data for all the scanned Ubuntu images. This data can be queried for further analysis and visualization.


\section{Findings and Results}

The dependency data extracted from a typical OS is a rich source of information on the high-level system architecture. In contrast to \textit{planned layer} of architecture, this layer refers to the \textit{unwritten} architectural aspects that emerge as a result of combining a multitude of independently-developed software packages. Coupled with temporal updates, this data can serve as a basis for a deeper system evolution trends analysis. For instance, long-term trends such as libraries gaining or losing popularity or executable complexity inflation may be detected. Predicting potential OS library or executable removal may help developers adjust the development plans. In addition, determining and removing unused libraries could be useful in optimizing disk space usage and reducing the attack surface.

Throughout the data collection conducted, we focused on three key aspects. Firstly, we investigated the OS-level dependency graph as a whole (RQ1). Secondly, we examined various aspects of complexity in binary dependencies determined through coupling analysis (RQ2). Lastly, we analyzed evolutionary trends in the OS dependency graph (RQ3).

\subsection{OS-wide Dependency Graph}

Analyzing the resulting SQLite database, which covers 84 Ubuntu images, revealed the following number of binaries, libraries and dependencies per image. We found that from Ubuntu 5.04 to 23.04 the number of binary executables ranged from \num{1519} to \num{2753} and the number of libraries ranged from \num{1683} to \num{3673}. In terms of dependencies detected, the numbers ranged from \num{18165} to \num[]{37641} in the images scanned. A total of \num{408364} binary and library files were processed to extract the dependencies, which returned almost 2 million dependencies. The total SQLite database size generated is over 83MB of raw dependency data.


We noticed that highly popular libraries such as (\texttt{libc}) make the graphs unreadable. Thus we implemented filtering out libraries from the sorted (by popularity) list of all the involved libraries. We observe that hiding the top 10-15 libraries increases the readability of the whole system graph. Notably, loosely coupled subsystems, such as the networking subsystem, become apparent. The libraries presented alongside the diagram also provide insight into the relative popularity of individual libraries within a system. 

We have observed that number of libraries imported but not present in the system varied from 20 (v5.04) to 8 (v23.04) with the highest number being 92 (v21.10b). As a consequence, the number of other libraries directly impacted by the missing dependencies varied from 4 (v17.10 and v17.10.1) to 27 (v13.04 and v9.04). Similarly, we see that the number of unused libraries (i.e., not imported by any other library or executable) ranged from 1301 (v5.04) to 1666 (v23.04). These numbers constitute a significant proportion of the total number of libraries included (around 77\% and 62\% respectively). Potential explanations for such a high number of unused libraries could be a) plugin-based applications that do not import libraries directly, b) "forgotten" legacy libraries and c) libraries shipped "just in case" for use by applications commonly installed at a later stage.

\vspace{-5pt}
\subsection{Dependencies Coupling Aspects}

Software dependencies represent the reliance of a given piece of code on external code. In practice, software developers only deal with a subset of the code required for an application to run. A graphics-oriented library may expose a simpler set of functions to developers, while relying on a multitude of other complex hardware-specific libraries to implement the advertised functionality. Thus, a complex and large code base is made to look simple from the developer's perspective.

This perception difference opens the possibility of measuring code coupling in direct and recursive ways. The direct coupling of an application reflects how many specific libraries a developer deals with explicitly. In contrast, recursive coupling takes all the underlying dependencies into consideration as well.

In addition, there is an inherent asymmetry in dependency tracking. Forward tracking from a given binary to all the required libraries is trivial, as this information is contained within the binary. Reverse tracking from a given library to determine all the binaries and libraries that require the specified library is complicated, as this information is not stored explicitly. Reverse tracking essentially reflects the popularity of a given library and requires scanning the whole file system to be calculated. Thus we developed functionality to measure the (i) direct coupling, (ii) total (recursive) coupling, and (iii) library popularity. 

\begin{figure}[h]
     \centering
     \begin{subfigure}[b]{0.49\textwidth}
         \centering      \includegraphics[width=\textwidth]{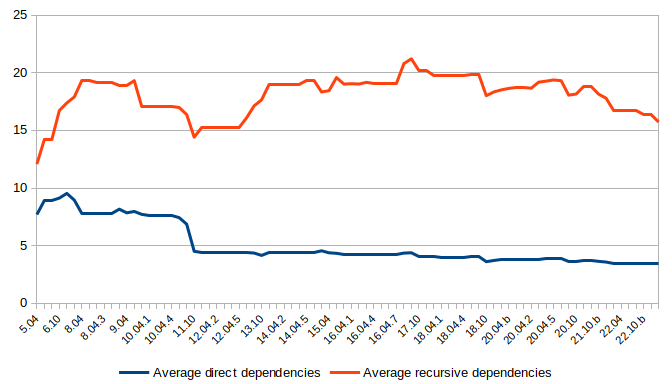}
    \caption{Average number of dependencies}
         \label{fig:averagedeps}
     \end{subfigure}
     \hfill
     \begin{subfigure}[b]{0.49\textwidth}
         \centering      \includegraphics[width=\textwidth]{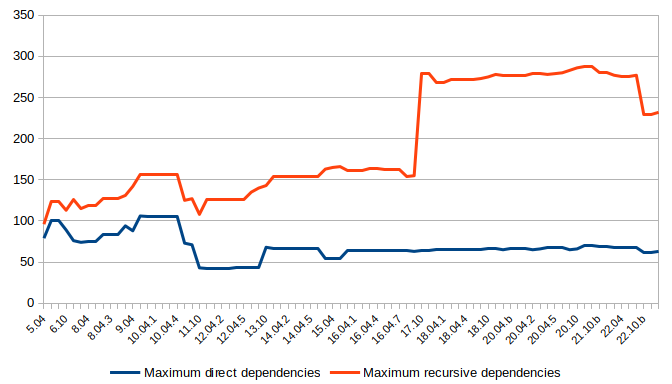}
    \caption{\small{Maximum number of dependencies}}
    \label{fig:totaldeps}
     \end{subfigure}  
     \hfill
    \caption{Direct and recursive dependencies}
    \label{fig:three graphs}
\end{figure}

Figures~\ref{fig:averagedeps} and \ref{fig:totaldeps} illustrate the changes in the average and maximum number of dependencies correspondingly. As can be seen from Figure \ref{fig:averagedeps}, whereas the average total number of dependencies largely stays the same, developer-facing complexity tends to decrease over time. This indicates that developers tend to re-arrange code within libraries to minimize the coupling they face directly. The large spike in Figure \ref{fig:totaldeps} is caused by the introduction of Gnome Shell in Ubuntu 17.10. We, therefore can conclude that while maintaining roughly the same external coupling, GNOME Shell has a complicated internal structure. Particularly, we found that \texttt{gnome-control-center} binary has the largest amount of dependencies. This is explained by the fact that the configuration tool needs to interact with most of the GNOME Shell subsystems.

A complementary aspect of dependency coupling is popularity. We define library popularity through the number of other libraries or executables that depend on it. In other words, damaging or removing more popular libraries would impact a larger number of executables in a system. In terms of popularity, the top 10 most used libraries (i.e. imported from other libraries and executables) in Ubuntu are: \texttt{libc (4397), libpthread (1438), libglib (1037), libgobject (945), libm (836), librt (719), libgthread (660), \\ libgmodule (658), libgtk-x11 (656), libdl (601)}. The numbers alongside the libraries refer to the number of uses  (i.e., library importing) averaged across all Ubuntu versions the library was present in.

We notice that 7 out of the top 10 directly-coupled libraries relate to various GNOME subsystems while the other 3 relate to the Evolution mail client. Interestingly, the most complex \texttt{ximian-connector-setup} executable with 100 direct dependencies was only present in two Ubuntu versions. This likely indicates that such high coupling was not tolerated, leading to the application removal.

Lastly, analyzing total coupling by taking recursive dependencies into account, we found the top 10 complex libraries and binaries:\texttt{empathy-call}(154),
 \texttt{evolution-alarm-notify}(156),  \texttt{gnome-control-center}(273), \texttt{gnome-todo}(155), \texttt{libvclplug\_gtk3lo}~(154), \texttt{smbd.x86\_64-linux-gnu}~(155), \texttt{libiradio}~(158),\\ \texttt{gnome-initial-setup}(169), \texttt{libgrilo}~(158),  
 \texttt{shotwell-publishing}~(164).
\vspace{-10pt}
\subsection{Dependency Graphs Evolutionary Trends}

\begin{figure}
    \centering
    \begin{subfigure}[b]{0.48\textwidth}
         \centering      
         \includegraphics[width=\textwidth]{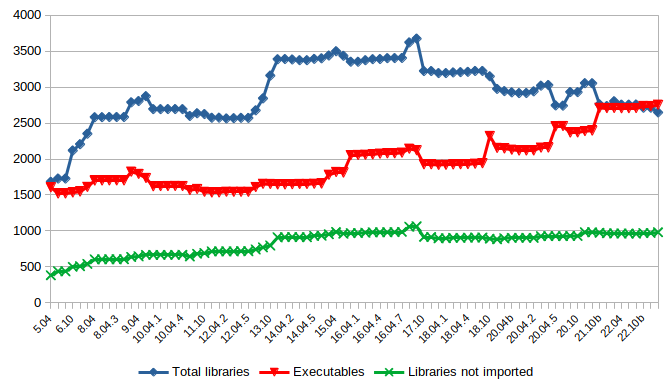}
    \caption{Libraries vs. executables evolution}
         \label{fig:libsevol}
     \end{subfigure}
     \hfill
     \begin{subfigure}[b]{0.48\textwidth}
         \centering      
         \includegraphics[width=\textwidth]{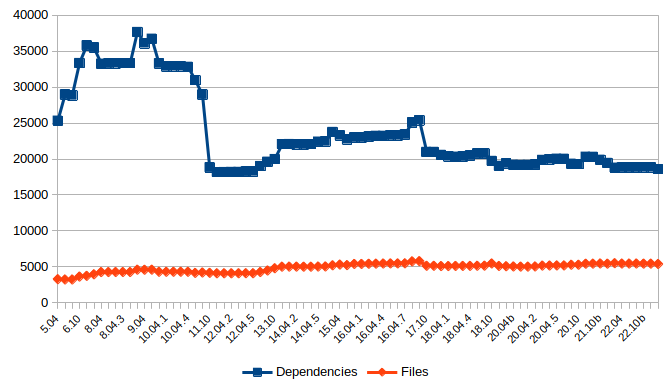}
    \caption{\small{Dependencies vs. Files evolution}}
    \label{fig:depsvsfiles}
     \end{subfigure}
     \hfill
    \caption{Overview of file-level evolutionary trends}
    \label{fig:filedepsstats}
    \vspace{-15pt}
\end{figure}

Running a large-scale analysis on a set of Linux distributions developed and released over 18 years revealed a number of shifts occurring in the domain. In constant efforts to attract users, Ubuntu is known for conducting experiments, such as introducing new large software packages as a replacement for existing ones. For instance, the significant dip in the number of dependencies on Figure \ref{fig:depsvsfiles} is explained by the replacement of GNOME 2 with Unity. On a longer scale it is also visible that despite limited local successes of such experiments, the overall trend indicates a slow growth of the number of files and dependencies. 

Interestingly, we also observed a significant amount of not explicitly required \texttt{.so} files are present in the system (Figure \ref{fig:libsevol}). In other words, up to 37\% of libraries physically located in the file systems were not mentioned in the import tables of any of the binaries or libraries. This likely indicates that such libraries are primarily used as plugins and could be loaded at run-time through dynamic directory scanning if necessary. Note that these conditional dependencies may be impossible to detect in advance due to the unpredictable nature of external factors. For instance, a user controlled application configuration can determine whether a given plugin library should be loaded at run time. The overall trend also hints that such a dynamic plugin-based approach gains popularity as the proportion of libraries not imported keeps steadily growing.  

Another observation discovered throughout our analysis relate to the longevity of the libraries and binaries in Ubuntu. Namely, while complex binaries are periodically removed in search of better alternatives, highly popular libraries tend to stay around. Once a popular library is introduced in a particular Ubuntu version, it is unlikely to be removed as such removal would impact all libraries and executables that rely on the library's existence. Even internal code reorganizations affecting highly popular libraries require extra care to maintain compatibility\footnote{https://developers.redhat.com/articles/2021/12/17/why-glibc-234-removed-libpthread}.

\vspace{-5pt}

\section{Discussion}

\subsection{Threats to Validity}

While we primarily focused on dependency-centric package management in Linux OS, other factors may explain some of the observations. Despite high popularity, packages might get removed from the system due to licensing, compatibility, security, or maintainability issues. Dependency analysis should, therefore, be coupled with change log analysis to verify and confirm the findings.

To enhance the external validity of our dependency analysis, we selected a highly popular Linux distribution. By including all of the available versions we expect our approach to be generalizable and applicable to a broader range of OSs. Widening the input data set on the time axis enabled the discovery of uncommon cases and long-term trends. 
Being well-maintained, Ubuntu served as a high-quality dataset. Legacy Ubuntu versions and their corresponding change logs were still available for download\footnote{Ubuntu wiki: Releases - https://wiki.ubuntu.com/Releases}. In contrast, Alpine (another popular Linux distribution) archives did not go far back in time. Moreover, the Alpine archives contained broken links for older versions, preventing image downloading. Similarly, while considering Debian systems, we discovered different and incompatible system image layouts which would complicate the analysis.

Primary threats to external validity are abrupt changes causing significant paradigm shifts, lower granularities skewing the results, and implicit dependencies.
\textit{Abrupt changes} may be introduced throughout evolution. Such changes introduce incompatibilities, forcing to amend the scanning process accordingly. Notable examples we observed include compression algorithm changes, folder hierarchy alterations, and transition from \texttt{RPATH} to \texttt{RUNPATH}. We noticed a different layout of binary files in the file system that required consideration due to the changes introduced in Ubuntu 19.04. Specifically, \texttt{/bin} and \texttt{/sbin} directories were converted to symbolic links to \texttt{/usr/bin} and \texttt{/usr/sbin} correspondingly\footnote{\url{https://lists.ubuntu.com/archives/ubuntu-devel-announce/2018-November/001253.html}}. Depending on whether 19.04 is being installed from scratch or on top of the previously installed version, the number of binaries may look like being suddenly doubled in version 19.04. We alleviated this problem by resolving symbolic links.

In addition to library dependencies stored in executable binary file import tables, other types of coupling occur in practice. For instance, network communication, special files like  Unix domain sockets, Inter-Process Communication (IPC) calls, message-oriented buses, and pipes provide various means of code interactions. Discovering such code coupling instances may not be possible in practice (e.g., new code fragments might be downloaded over a network). Taking into account these code coupling types may significantly skew our findings.

\vspace{-10pt}
\subsection{Challenges and Limitations}

The two primary technical challenges we encountered throughout our data collection and analysis are the large data set sizes and performance issues related to extracting dependencies at lower granularities.

As the distributed Ubuntu images are growing in size, so do the number of executable files and their individual sizes. This steady growth is observed over all Ubuntu versions analyzed. For example, within 18 years analyzed, the live Ubuntu image size grew from 600MB (version 5.04) to 3.7GB (version 23.04). Likewise, the number of executable files experienced a 70\% increase in size (\num{1605} in 5.04, \num{2753} in 23.04).

Through practical experiments, we established that restricting the dependency granularity is crucial to achieving acceptable processing speed as lower granularity dependency extraction incurs large overheads. Disassembling executable binaries to identify individual third-party library function calls slows the dependency extraction and incurs significant memory overheads. For instance, we have observed cases of over-disassembly and analysis of a single executable taking 40 minutes on an average laptop-class CPU. Thus, while technically possible and potentially interesting to gain further insights, lower-level granularity analysis is out of reach for real-time applications we initially aimed for. At this stage, we restricted the analysis to the file level only.

\section{Related Work}

The prior work primarily revolves around two aspects, (i) diverse conceptual complexity metrics definitions and (ii) dependency extraction and analysis.

Various types of software complexity metrics have been widely studied in the literature \cite{StaticDependency}. Some studies have focused on metrics that are useful in source code analysis but are not easily applicable in binary code analysis \cite{SoftwareMetrics} \cite{SoftwareMetricsSurvey} \cite{ComplexityComparison}. Others have discussed the deficiency of methods to obtain global dependency knowledge and the difficulty in visualizing the resulting graphs \cite{PackageDependency_2015}. The use of software complexity metrics to detect vulnerabilities has also been investigated, with some studies proposing dependency-oriented and execution-time complexities \cite{InitialComplexity}. Dependency extraction aspects and challenges have also been explored, with some studies focusing on specific languages or ecosystems \cite{TopologyAnalysis} \cite{EmpiricalComp}.

Package management and dependency validation have been popular research topics, with a set of studies proposing methods to address issues arising from package evolution (e.g., splitting into multiple different packages) \cite{PackageConflict} \cite{{DebianLinux}} \cite{LinuxPackage_IEEE}. User questions related to package management, such as calculating the consequences of removing or modifying a package, have also been explored \cite{LinuxPackageVis} \cite{EvolutionPackageDepen}.
Efficient package management tools and query languages have been proposed, including tools for efficient package management and relations lookup \cite{LinuxQuality}. However, similar to software complexity metrics research efforts, multiple studies have focused only on source-level rather than binary dependencies \cite{RecoverDependency} \cite{AutoDepen}. 

In efforts to resolve binary compatibility issues, some works have investigated relying on version ranges rather than minimum version requirements \cite{DepOwl}. Unfortunately, the large downside of the proposed approach is the requirement of debug symbols availability, which is rare in commercial software. An interesting use of dependency extraction has been proposed for Windows executables for malware detection \cite{DLLMiner}. Taking the notion of the extent of a dependency into account enables detecting and eliminating insignificant dependencies \cite{SurviveDependency}. 

Overall, it should be noted that dependency related studies primarily focus on source code dependency analysis and package-level relations\cite{interPackage} \cite{LinuxDis} and do not typically examine software package evolution over time. We, therefore, conclude that a more precise file-based dependency extraction is an  under researched area that might benefit from providing better structural visibility for large-scale systems comprising multiple independently developed packages. We also see that understanding software evolution is essential for maintaining software, ensuring compatibility, and improving security. Having this understanding aids developers in making informed decisions about updates and maintenance, ensures software remains compatible with other systems, and reduces the risk of security issues. Additionally, understanding software evolution can lead to new innovations and improvements in software design and development.

\vspace{-10pt}
\section{Conclusion and Future Work}
\vspace{-10pt}

In this study, we introduce automated extraction of dependency graphs for a whole system at the executable files level (as opposed to manually maintained traditional package-level dependency graphs). The resulting system-wide dependency graph provides a high-level view of the OS architecture emerging from interactions between the different subsystems and user packages. In addition, this study enabled the discovery of general high-level trends/common patterns in Ubuntu Linux architecture evolution over time. 

We also differentiate between developer-facing complexity (defined through direct dependency coupling) and overall system complexity (defined through recursive dependency coupling). The motivation behind such a separation is that developers typically deal with third-party libraries without having full visibility of the back-end side of the libraries. In other words, a developer may include one library, while the library itself can have a complicated graph of dependencies not directly visible to the developer. These invisible dependencies may cause software bloating and increase the attack surface. 
We believe the findings of this study will provide useful insights for software developers and OS maintainers in terms of gaining a holistic quantitative understanding of inter-package architecture management that would be useful, for example, in optimizing disk space and improving system maintainability.


We have identified two main directions for future research lines. Specifically, expanding the dependency extraction approach to a \textit{wider set of platforms} to support and \textit{more types of dependencies} to extract.
For future research, we aim to perform Windows-based analysis and implement support for other levels of granularity, such as individual function dependencies. Also, in contrast to the convenient, holistic file system structure used in live editions, non-live distribution variants are composed of multiple compressed packages, complicating the dependency extraction and analysis. Implementing analysis for such non-live distributions could be a potential future research line.

As opposed to fixed library imports, code fragments interacting through various communication channels are loosely coupled. Such non-obvious dependencies are not trivial to detect. For instance, changing code on one side of a UNIX pipe may negatively affect the results of the next program in the pipeline. Furthermore, such dependencies may not be predefined in advance and are only required intermittently while being completely unnoticeable most of the time. We believe that comprehensive and accurate detection of such concealed dependencies would greatly enhance the overall system architecture, evolution, and run-time operation understanding and visibility and enable early detection of potential compatibility breaks caused by code modifications.

\vspace{-5pt}
\section*{Acknowledgment}
The work has been partially supported by the Cyber Security Research Centre Limited whose activities are partially funded by the Australian Government’s Cooperative Research Centres Programme.
\vspace{-10pt}

\section*{Data Availability}
\vspace{-5pt}
As the current project is funded by industry partners, we are unable to publish the source code at this stage. However, aiming to increase transparency and reproducibility in research, we have made the obtained dataset available for public access~\cite{DataLink}. Researchers and interested parties can access the dataset and utilize it to replicate or build upon our findings.

\end{document}